\documentclass[fleqn,10pt]{wlscirep}
\usepackage[utf8]{inputenc}
\usepackage[T1]{fontenc}

\usepackage{graphicx}  
\usepackage{sidecap}
\usepackage{color, colortbl}
\definecolor{Gray}{gray}{0.9}

\title{Resonant plasmonic detection of terahertz radiation in  field-effect transistors  with the graphene channel and the black-As$_x$P$_{1-x}$   gate layer  
}

\author[1,*]{V.~Ryzhii}
\author[1]{C.~Tang} 
\author[1]{T.~Otsuji} 
\author[2]{M.~Ryzhii}
\author[3]{V.~Mitin}
\author[4]{M.~S.~Shur}

\affil[1]{Research Institute of Electrical Communication, Tohoku University, Sendai 980-8577, Japan}
\affil[2]{Department of Computer Science and Engineering, University of Aizu, Aizu-Wakamatsu 965-8580, Japan}
\affil[3]{Department of Electrical Engineering, University at Buffalo, SUNY, Buffalo, NY 14260, USA}
\affil[4]{Department of Electrical, Computer, and Systems Engineering, Rensselaer Polytechnic Institute, Troy, NY 12180, USA}
\affil[*]{v-ryzhii @ riec.tohoku.ac.jp}

\begin{abstract} 
\normalsize We propose the terahertz (THz) detectors based on field-effect transistors (FETs) with the graphene channel (GC)  and the black-Arsenic (b-As) black-Phosphorus (b-P),  or black-Arsenic-Phosphorus
(b-As$_x$P$_{1-x}$) gate barrier layer.
The operation of the GC-FET detectors is associated with the carrier heating in the GC by the THz electric field
resonantly excited by incoming radiation leading to an increase in the rectified current 
between the channel and the gate over the b-As$_x$P$_{1-x}$ energy barrier layer (BLs). 
The specific feature of the GC-FETs under consideration is relatively low energy BLs
and the possibility to optimize the device characteristics by choosing the barriers containing a
necessary number of the b-As$_x$P$_{1-x}$ atomic layers and a proper gate voltage.
The excitation of the plasma oscillations in the GC-FETs leads
to the resonant reinforcement of the carrier heating and the enhancement of the detector responsivity. 
The room temperature responsivity can exceed the values  of $10^3$~A/W.
The speed of the GC-FET detector's response to the modulated THz radiation is determined by the processes of  carrier heating.
As shown, the modulation frequency can be in the range of several GHz at room temperatures.
\end{abstract} 
\begin{document}

\flushbottom
\maketitle
%
%
\thispagestyle{empty}


\section*{Introduction}

The emergence of the black-Phosphorus (b-P), black-Arsenic (b-As), and the compounds of these materials (b-AsP), with the energy gap $\Delta_{BL}$ varying from 0.15 to 1.2 eV
(see, for example, \cite{1,2,3,4,5,6,7,8,9,10,11,12,13,14,15,16,17,18,19,20,21})  opens new prospects for the creation of different electronic, optoelectronic, and terahertz (THz) devices.  The combination of the GLs with the b-As$_x$P$_{1-x}$ layers with
graphene~\cite{22,23,24,25,26,27,28} can be particularly beneficial for the creation of  novel devices, including MIR/FIR/THz interband photodetectors. In this paper, we propose and evaluate  the THz detectors
akin to the field-effect transistors (FETs) with the graphene channel (GC) and b-As$_x$P$_{1-x}$ gate barrier layer (BL).
The operation of such GC-FETs is associated with  the carrier heating in the GC
by incoming THz radiation (see, for example,~\cite{29}) leading to an increase of the thermionic GC-gate current.
This implies that the GC-FETs could operate as  hot carrier bolometric detectors.
The main features of the proposed THz detectors are as follows: 
(a) the b-As$_x$P$_{1-x}$ BL
provides the possibility to choose the desirable BL height (and, hence, optimize the device characteristics)  by varying the number of the atomic layers and/or the molar fraction  of As~\cite{3,4,5,6},
(b) the GC exhibits a room temperature elevated carrier energy    
relaxation time~\cite{30,31,32,33,34}, which promotes  high detector responsivities and detectivities, and
(c) the plasmonic (PL) properties of the GC-FET~\cite{35,36,37} can enable the detector resonance response to the THz radiation at the frequencies close to the GC frequencies.

\begin{figure*}\centering
\includegraphics[width=9.0cm]{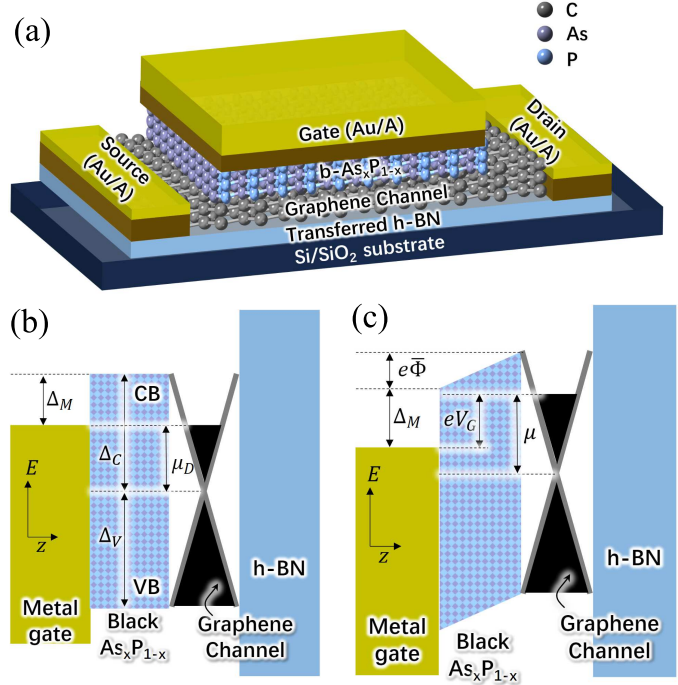}
\caption{
(a) Sketch of the GC-FET detector structure 
and  the  band diagrams of the GC-FET with $\Delta_C -\mu_D = \Delta_M$ at (b) $V_G =0$ (BL flat band condition) and ${\overline j} = 0$,
 and (c) $V_G > 0$ with $\Delta_C - \mu < \Delta_M$ and $\Delta_V+\mu > \Delta_M$, hence 
 the thermionic electron current density ${\overline j} > 0$  (the hole current is negligible). 
} 
\label{F1}
\end{figure*}

\begin{table}[t]
\centering
\vspace{2 mm}
\begin{tabular}{|r|c|c|c|c|c|c|c|}
\hline
\cellcolor{Gray}
&\cellcolor{Gray} $x$&\cellcolor{Gray}  Metal &\cellcolor{Gray}  $\Delta_C$(meV)&\cellcolor{Gray} $\Delta_V$(meV)&\cellcolor{Gray} $\Delta_M$(meV)&\cellcolor{Gray} $\mu_D$~(meV) &\cellcolor{Gray}$\Sigma_D^{=}$~(10$^{12}$cm$^{-2}$)\\ 
\hline
Sample 1a&	0 & 	Al&	225&	125&	85& 140  & 1.60\\ 
Sample 2a&	1.0&	Al&	190&	90&	     50&140   &1.60\\
\hline
Sample 1b&	0	&   Ti&	225&	125&	105& 120 &1.17\\
Sample 2b&	1.0&	Ti&	190&	90&	    70& 120 &1.17\\
\hline
Sample 1c&	0&	    In&	225&	125&	135& 90 &0.66\\
Sample 2c&	0.5&	In&	230&	130&	130& 100 &0.81\\
\hline
Sample 1d&	0&  	Mn&	225&	125&	125&100&0.81\\
Sample 2d&	0.5&	Mn&	230&	130&	120&110 &0.89\\
\hline
\end{tabular}
\caption{\label{table} Metal gate/b-As$_x$P$_{1-x}$/GC barrier layer parameters, $N = 20$, $W = 10$~nm~\cite{11,12,13,14,15,16,17,18,19,20,21}.} 
\end{table}

\section*{Device structure}
Figure~1(a) shows the GC-FET detector  structure (with the number of atomic layers in the BL $N=20$) and the related band diagrams for the assumed band alignment.
 For definiteness, we consider the GC-FET structures with the n-type GC, in which 
  the thermionic current between the GC and the gate is associated with the electrons overcoming
the BL barrier in the conduction band. If 
 $\Delta_V \geq \Delta_C$, where
$\Delta_C$ and $\Delta_V$ are  the band offsets between the BL conduction and valence band's edges and the Dirac point in the GC (so that $\Delta_C + \Delta_V$ is the energy gap of the BL), the electron current exceeds the hole current when
 the electron Fermi energy, $\mu_D$, in the GC is sufficiently large, so that $\Delta_C - \mu_D \lesssim \Delta_M$. Here
  $\Delta_M$ is the difference between the BL and GC work functions and $\mu_D \simeq \hbar\,v_W\sqrt{\pi\Sigma_D^{=}}$ is the equilibrium 
  value of the electron Fermi energy  counted from the Dirac point at $V_G = 0$, $v_W \simeq 10^8$~cm/s is the characteristic electron velocity in GCs, and $\hbar$ is the reduced Planck constant.
 
Figures 1(b) and 1(c) show  the band diagrams  for $\Delta_C - \mu_D = \Delta_M$
at zero  gate bias ($V_G =0$ with the GC donor density $\Sigma_D^{=}$  corresponding  to the BL flat band)
and under the gate bias ($V_G > 0$), respectively.
 We focus on the GC-FETs with n-doped GC, in which the above equality is met, 
i.e.,  
 assuming that in the absence of the bias gate voltage ($V_G = 0$), the BL bottom of the conduction band and the top of the valence band are flat.

  Table I lists  examples of possible combinations of the
 b-As$_x$P$_{1-x}$ barriers and the metal gate materials.
The pertinent parameters were taken from~\cite{11,12,13,14,15,16,17,18,19,20,21}.

 The results obtained below can be also used for devices with relatively large $\Delta_C$, considering  the hole transport instead of the electron one.

 The local voltage drop across the BL
$e\Phi = \Delta_C - \Delta_M - \mu + e(V_G + \varphi) =
\mu_D-\mu + e(V_G + \varphi)
$, 
where  $V_G$ is the applied DC bias gate voltage,  $\varphi = \varphi(x,t)$ is the GC potential  local value, and $\mu$ is the net electron Fermi energy in the GC.
At the GC edges, $\varphi(\pm L, t) = \pm \frac{1}{2}\delta  V_{\omega}\exp(-i\omega t)$ or $\varphi(\pm L, t) =  \delta  V_{\omega}\exp(-i\omega t)$
 for the asymmetric (a) and symmetric (s) THz radiation input, respectively,
with
 $\delta V_{\omega}$ and $\omega$
being the amplitude and the frequency 
of the THz radiation received by an antenna.
The asymmetric THz  radiation input corresponds to the design when  the antenna leads are connected to the GC-FET side (source and drain) contact pads. In the case of the symmetric input, one of the antenna leads is contacted to the gate, whereas the second one to both side contacts.  

The variation of the potential difference between the side source and drain contacts (in the case of the symmetrical input) and between the side contacts
and the gate leads to the transient  electron current along the GC and the transient  variation
of the self-consistent electron density, i.e., to the excitation of the plasmonic oscillations. In the case of the asymmetric input, the electron current along the GC exists even at very small signal frequency. As a results the electron heating
by the incoming signals takes place at such frequencies as well. 
In contrast, in the case of the symmetrical input, slow variations of the 
side contacts potential with respect to the gate potential create a very weak lateral electron current not heating the GC electron system. This leads to  marked distinctions of the response in the range of low frequencies (which is demonstrated below).


 The BL energy gap $\Delta_G$ and the dielectric constant $\kappa_G$ depend on the transverse electric field $\Phi/W$.
 Accounting for this, one can   set $\Delta_C = \eta \Delta_G[1 - (\Phi/WE_G)^2]$, 
 and $\kappa = \kappa_G/[1 - (\Phi/WE_G)^2]$,
 where $ \Delta_G$ and $\kappa$ are the BL energy gap and the dielectric constant in the absence of the transverse electric field,  $E_G$ is the characteristic electric field, $W$ is the BL thickness (see, for example,~\cite{38,39,40,41}), and $\eta = \Delta_C/(\Delta_C+\Delta_V)< 1$ the fraction of the
BL height related to the conduction band.
For the b-P BL with $W = 10$~nm  (the number of the atomic layers $N = 20$),
$E_G \simeq 0.7 - 0.8\simeq $~V/nm. This implies that the effect of the transverse electric field on
$\Delta_C$, $\Delta_G$, and $\kappa_G$ markedly reveals at sufficiently high gate voltages when $\Phi \gtrsim 1$~V). However, such a voltage range is beyond our present consideration.
Considering the GC-FETs with a sufficiently thick  b-As$_x$P$_{1-x}$ BL at moderate gate voltages,
  we disregard the carrier
tunneling across this layer. This implies that  the GC-gate current is
associated with the sufficiently energetic electrons overcoming the BL, i.e., it is of thermionic origin.

\section*{Equations of the model}

\subsection*{Thermionic DC and AC}

At not-too-small electron densities in GCs, the characteristic time of the electron-electron collisions $\tau_{ee}$ is shorter than the pertinent times associated with the optical phonons $\tau_0$, acoustic phonons $\tau_{ac}$, and impurities  $\tau_i$, respectively.
This implies that the electron distribution function is close to the Fermi distribution function
$f(\varepsilon) = [\exp(\varepsilon - \mu)/T +1]^{-1}$, characterized by the effective electron
temperature $T$ generally different from the lattice (thermostat) temperature $T_0$ (in the energy units) and the electron Fermi energy $\mu$. Hence,  at $\varepsilon > \mu$,  $f(\varepsilon) \simeq \exp[(\mu - \varepsilon)/T]$. However, in the energy range $\varepsilon > \Delta_C$, the electron escape over the BL can markedly decrease $f(\varepsilon)$.
To account for this effect, in the range in question, one can set $f(\varepsilon) \simeq \xi\exp[(\mu - \varepsilon)/T]$, where $\xi = \tau_{\bot}/(\tau_{ee} + \tau_{\bot})$ with $\tau_{\bot}$ being
   the electron try-to-escape time.

 Considering that the height of the potential barrier for the electrons in the GC and in the metal gate are equal to  $\Delta_C - \mu$ and $\Delta_M + e(V_G + \varphi)$, respectively, 
the density of the  thermionic electron  current  can be presented as

\begin{eqnarray}\label{eq1}
j \simeq  j^{max}
\biggl[\exp\biggl(\frac{\mu   - \Delta_C}{T}\biggl) - \exp\biggl(-\frac{\Delta_M + e \Phi}{T_0}\biggr)\biggr].
\end{eqnarray}  
Here $j^{max}= e\Sigma/\tau_{\bot}$ is the characteristic (maximum)  GC-gate DC  density, $\Sigma$ is the electron density  in the GC induced by the donors and  gate voltage, and
 $e =|e|$ is the electron charge. One can assume that $\tau_{\bot}$ is determined by the momentum relaxation time, associated with  the quasi-elastic scattering of the high-energy electrons, i.e., with acoustic phonons  (in sufficiently perfect GCs). Due to this, 
it is natural to assume that $\tau_{\bot} > \tau_{ac} \gg \tau_{ee}$.
The Fermi energy $\mu$ is determined by both the GC doping and the gate voltage.

Equation~(1) leads to the following expressions for the thermionic DC density ${\overline j}$, corresponding to the DC temperature $\overline T$: 

\begin{eqnarray}\label{eq2}
 {\overline j}=j^{max}\biggl[\exp\biggl(\frac{\mu   - \Delta_C}{{\overline T}}\biggl) -\exp\biggl(\frac{\mu -\Delta_C -eV_G}{T_0}\biggr)\biggr].
\end{eqnarray} 
Due to the dependence of $\mu$ on  $V_G$, Eq.~(2) provides the GC-gate I-V characteristics.
Since ${\overline T}$ also depends on $V_G$ (because of the electron heating in the GC by the lateral DC), the latter dependence can somewhat contribute to the GC-FET characteristics as well.
 
At sufficiently high GC lateral conductivity in the situations under consideration (large $\Sigma$
and $\mu$),
the DC potential and the DC effective temperature nonuniformity along the GC are weak (${\overline T}\simeq const$). This implies that we disregard the possible DC crowding. A high electron thermal conductivity additionally suppresses  the above nonuniformity.

The AC variation $\delta j_{\omega}$  due to the potential oscillations  leading to the electron
heating is given by
 
\begin{eqnarray}\label{eq3}
\delta j_{\omega} = j^{max}\frac{\delta T_{\omega}}{ {\overline T}}
\frac{(\Delta_C -\mu)}{ {\overline T}}\exp\biggl(\frac{\mu   -  \Delta_C}{ {\overline T}}\biggr)
  \end{eqnarray} 
Here we omitted the term containing the factor $ (e\delta \varphi_{\omega}/T_0 )^2/2$
with  $\delta \varphi_{\omega}$ being  the GC potential ac component.
In this case, 
the quantity $\delta j_{\omega}$, given by Eq.(3), does not depend explicitly on the AC variations of the GC potential
(only via the effective temperature variation $\delta T_{\omega}$).
This is due to a specific shape of the energy barrier for the electrons in the GC [see Fig.~1(c)].

\subsection*{Rectified  current and effective carrier temperature}

The incoming THz radiation results in variations of the potential in the GC. This leads to
 extra electron heating and the variation of the electron temperature $\delta T = T - {\overline T}$.
According to Eq.~(3), the variation of the net gate current associated with the effect of the incoming THz radiation averaged over its period  (rectified photocurrent)
 is given by

\begin{eqnarray}\label{eq4}
<\overline{\delta J_{\omega}}> = J^{max}{\mathcal F}(V_G) 
\frac{<\overline{\delta T_{\omega}}>}{{\overline T}}.
\end{eqnarray} 
Here $J^{max} = 2LHj^{max}$, and  $2L$ and $H$ are the GC length and width,

\begin{eqnarray}\label{eq5}
{\mathcal F}(V_G) =  \frac{(\Delta_C -\mu)}{{\overline T}}\exp\biggl(\frac{\mu   -  \Delta_C}{{\overline T}}\biggr)
= \frac{[\Delta_M -(\mu-\mu_D)]}{{\overline T}}\exp\biggl[\frac{(\mu-\mu_D)   -  \Delta_M}{{\overline T}}\biggr]
\end{eqnarray} 
is the barrier factor, and
the symbols $<...>$  and $\overline{<...>}$ denote the averaging over the signal period $2\pi/\omega$ and the length of the GC, respectively,
with 
\begin{eqnarray}\label{eq6}
<\overline{\delta T_{\omega}}> = \frac{1}{2L}\int_{-L}^Ldx<\delta T_{\omega}>.
\end{eqnarray}

The dependence of the factor ${\mathcal F}(V_G)$ on the gate voltage  as associated with 
the voltage dependence of  the electron Fermi energy (see below).

 The  effective electron temperature $T$ is determined by the  balance of the electron energy
 transfer to the lattice and the energy provided by the electric field along the GL.
 At room temperature, the emission and absorption of the optical phonons by the electrons in GLs
 can be considered as a main mechanism of  electron energy relaxation.
In this case,
the power transferring from the electrons in the GC to the optical phonons due to the intraband transitions
is ~\cite{30,31,32,33}
\begin{eqnarray}\label{eq7}
P_0^{intra} = \hbar\omega_0 R_0^{intra}.
\end{eqnarray} 
Here

\begin{eqnarray}\label{eq8}
 R_0^{intra} = R_0 \frac{\hbar\omega_0\mu^2}{T_0^3}\biggl[\biggl(1 + \frac{1}{{\mathcal N}_0}\biggr)\exp\biggl(-\frac{\hbar\omega_0}{T}\biggr)- 1\biggr],
\end{eqnarray} 
$\hbar\omega_0 \sim 200$~meV is the optical phonon energy, ${\mathcal N}_0 = [\exp(\hbar\omega_0/T_0)-1]^{-1} \simeq \exp(-\hbar\omega_0/T_0)$,  $R_0$ is the characteristic rate of the interband absorption of  optical phonons,  and $T_0$ is the lattice temperature.
At moderate THz power, the effective electron temperature $T$ is close to the optical phonon temperature $T_0$, and Eq.~(8) yields for  $R_0^{intra}$:

\begin{eqnarray}\label{eq9}
  R_0^{intra} \simeq R_0 \frac{\hbar\omega_0\mu^2}{T_0^3}\biggl(\frac{1}{T_0} - \frac{1}{T}\biggr).
\end{eqnarray} 

Equalizing $ R_0^{intra}$ given by Eq.~(9) and the Joule power associated with the AC in the GC, for
 the THz range of frequencies (in which one can assume  $\omega \gg 1/\tau_{\varepsilon}$),
 we arrive at the following energy balance equation: 

\begin{eqnarray}\label{eq10}
 \frac{<\overline{\delta T_{\omega}}>}{\tau_{\varepsilon}} = 
 \frac{{\rm Re}~\sigma_{\omega}}{2\Sigma_0L}
 \int_{-L}^{L}dx\biggl|\frac{d\varphi_{\omega}}{dx}\biggr|^2.
\end{eqnarray} 
Here Re~$\sigma_{\omega} = \sigma_0\nu^2/(\nu^2+\omega^2)$ is the real part of the GC Drude conductivity,
$\sigma_0 = e^2\mu/\pi\hbar^2\nu$ is its DC value, $\nu$ is the frequency of the electron collisions
on impurities, acoustic phonons, as well as due to the carrier viscosity~(see,~\cite{42} and the references therein). Accounting for the deviation of the optical phonon temperature $T_0$ from the lattice temperature $T_l$, the carrier energy relaxation time   $\tau_{\varepsilon}$ 
associated with the interaction with optical phonons
is estimated as~\cite{32}
 $\tau_{\varepsilon} = \tau_0 (1 + \xi_0)(T_l/\hbar\omega_0)^2\exp(\hbar\omega_0/T_l) \simeq \tau_0 (1 + \xi_0)(T_0/\hbar\omega_0)^2\exp(\hbar\omega_0/T_l)$, where
  $\tau_0$ is the characteristic  time of the spontaneous optical phonon intraband emission
by the electrons and $\xi_0 = \tau_0^{decay}/\tau_0$, and $\tau_0^{decay}$ is the decay time of optical phonons in GCs.

\subsection*{Plasmonic oscillations factor}

The description of the spatio-temporal oscillations of the electron density and the self-consistent electric field, i.e.,
the plasmonic oscillations  in the GLs (see, for example,~\cite{32,33,34,35,36,37}) forced by the incoming THz signals can be reduced to
a  differential equation for the AC potential of the gated GC filled by the electrons (followed from a hydrodynamic electron transport model equations~\cite{43,44,45} coupled with the Poisson equation), $\delta \varphi_{\omega}(x)$ :

\begin{eqnarray}\label{eq11}
 \frac{d^2\delta \varphi_{\omega}}{dx^2} 
 + \frac{\omega(\omega+i\nu)}{s^2}\delta \varphi_{\omega} =0,
\end{eqnarray} 
 supplemented by the following boundary conditions:

\begin{eqnarray}\label{eq12}
 \delta \varphi_{\omega}|_{x= \pm L}| =\pm \frac{\delta V_{\omega}}{2}\exp(-i\omega t), \qquad  \delta \varphi_{\omega}^s|_{x= \pm L}| = \delta V_{\omega}\exp(-i\omega t).
\end{eqnarray} 
Here  $s = \sqrt{4\,e^2\mu\,w/\kappa \hbar^2}$  is the plasma-wave velocity in the gated GC. 

The above equations  yield 
the following formula for the AC potential along the GC

\begin{eqnarray}\label{eq13}
\delta\varphi_{\omega}^a = \frac{\delta V_{\omega}}{2}\frac{\sin(\gamma_{\omega}x/L)}{\sin\gamma_{\omega}}, \qquad \delta\varphi_{\omega}^s = \delta V_{\omega}\frac{\cos(\gamma_{\omega}x/L)}{\cos\gamma_{\omega}}.
\end{eqnarray} 
Here 

\begin{eqnarray}\label{eq14}
\gamma_{\omega} = \pi\frac{\sqrt{\omega(\omega+i\nu)}}{\Omega}, 
\qquad
\Omega  = \sqrt{\frac{4\pi^2\,e^2\mu\,W}{\kappa \hbar^2L^2}} 
\end{eqnarray} 
are the normalized wavenumber  and 
 the characteristic  frequency of the plasmonic oscillations of the
  electron system in the GC-FET under consideration.

The AC electric field along the GC is equal to

\begin{eqnarray}\label{eq15}
\frac{d\varphi_{\omega}^a}{dx} = \frac{\delta V_{\omega}}{2}
\frac{\gamma_{\omega}}{L}\frac{\cos(\gamma_{\omega}x/L)}{\sin\gamma_{\omega}},
\qquad
\frac{d\varphi_{\omega}^s}{dx} = -\delta V_{\omega}
\frac{\gamma_{\omega}}{L}\frac{\sin(\gamma_{\omega}x/L)}{\cos\gamma_{\omega}}
\end{eqnarray}
that, accounting for Eq.~(12), yields

\begin{eqnarray}\label{eq16}
\frac{<\overline{ \delta T_{\omega}}>^{a,s}}{\tau_{\varepsilon}}
 = \biggl|\frac{\delta V_{\omega}}{2}\biggr|^2
 \frac{\sigma_0}{{\overline\Sigma}L^2}{\mathcal P}_{\omega}^{a,s}.
\end{eqnarray}
Here

\begin{eqnarray}\label{eq17}
{\mathcal P}^a_{\omega} =
 \frac{\nu^2}{(\nu^2+\omega^2)}
\int_0^1d\zeta\biggl|\frac{\gamma_{\omega}\cos(\gamma_{\omega}\zeta)}{\sin\gamma_{\omega}}\biggr|^2, \qquad  
{\mathcal P}^S_{\omega} =
 \frac{\nu^2}{(\nu^2+\omega^2)}
\int_0^1d\zeta\biggl|\frac{\gamma_{\omega}\sin(\gamma_{\omega}\zeta)}{\cos\gamma_{\omega}}\biggr|^2
\end{eqnarray}
are the plasmonic factors,  which can be also presented as

\begin{eqnarray}\label{eq18}
{\mathcal P}^a_{\omega}
\simeq \biggl(\frac{\pi\nu}{\Omega}\biggr)^2
\frac{\omega}{\sqrt{(\nu^2+\omega^2)}}\frac{ P^a_{\omega}}{|\sin\gamma_{\omega}|^2}, \qquad 
{\mathcal P}^s_{\omega}
\simeq \biggl(\frac{\pi\nu}{\Omega}\biggr)^2
\frac{\omega}{\sqrt{(\nu^2+\omega^2)}}\frac{ P^s_{\omega}}{|\cos\gamma_{\omega}|^2},
\end{eqnarray}
with 
$P^a_{\omega} = \int_0^1d\zeta|\cos(\gamma_{\omega}\zeta)|^2$ and
$P^s_{\omega} = \int_0^1d\zeta|\sin(\gamma_{\omega}\zeta)|^2$ 
being  functions of the order of unity oscillating with  the frequency. 
If $\omega \ll \Omega$ ($\gamma_{\omega}$ tends to zero), Eqs.~(17) and (18)  yield ${\mathcal P}^a_{\omega} \simeq 1$ and  ${\mathcal P}^s_{\omega} \simeq 0$.

Combining Eqs.~(4), (6), and (16), we obtain  

\begin{eqnarray}\label{eq19}
\frac{<\overline {\delta J_{\omega}}>^{a,s}}{J^{max}} = \biggl|\frac{\delta V_{\omega}}{2}\biggr|^2\frac{\sigma_0\tau_{\varepsilon}}{{\overline \Sigma}L^2} {\mathcal F}(V_G)\,{\mathcal P}^{a,s}_{\omega}.
\end{eqnarray}

The detector response depends on the antenna type (see, for example,\cite{46,47}).
Using an antenna specially desined for the THz range could substantially increase the collected power~\cite{47}.
Here we define  the  GC-FET  detector current responsivity (in the A/W units)
and its voltage responsivity  (in the V/W units) as

\begin{eqnarray}\label{eq20}
{\mathcal R}_{\omega} = \frac{<\overline {\delta J_{\omega}}>^{a,s}}{ S_{\omega}},\qquad {\mathcal R}_{\omega}^V = \frac{<\overline {\delta J_{\omega}}>^{a,s}}{S_{\omega}} \rho,
\end{eqnarray}
respectively. Here $S_{\omega}$
is the THz power collected by an antenna
 and
$\rho = 2L/H\sigma_0$ is the GC DC resistance (for the case of  load resistance equal to the GC resistance).
This collected power
 is estimated as $S_{\omega} = I_{\omega}A_{\omega}$, where
$I_{\omega}$ is the intensity of the impinging radiation and $A_{\omega} = \lambda_{\omega}^2g/4\pi$ is the antenna aperture~\cite{46}, $\lambda_{\omega}$ is the radiation wavelength, and $g$
is the antenna gain. 
Consideringm as an example, the half-wavelength dipole antenna, for which $|\delta V_{\omega}|^2 \simeq I_{\omega} (8\pi/c)(\lambda_{\omega}/\pi)^2$, where $c$ is the speed of light in vacuum, 
we obtain 
$|\delta V_{\omega}|^2 = 32S_{\omega}/gc$.

 Accounting for Eqs.~ (18) and (19), we obtain

\begin{eqnarray}\label{eq21}
{\mathcal R}_{\omega} = \frac{32}{g c}\frac{<\overline {\delta J_{\omega}}>^{a,s}}{|\delta V_{\omega}|^2}, \qquad
 {\mathcal R}_{\omega}^V = \frac{32}{gc}\frac{<\overline {\delta J_{\omega}}>^{a,s}}{|\delta V_{\omega}|^2}\,\rho.
\end{eqnarray}
The latter equations 
 yield

\begin{eqnarray}\label{eq22}
{\mathcal R}_{\omega}= {\mathcal R}_0 {\mathcal F}(V_G){\mathcal P}^{a,s}_{\omega},\qquad
{\mathcal R}_{\omega}^V = {\mathcal R_0}^V {\mathcal F}(V_G)\,{\mathcal P}_{\omega}^{a,s},
\end{eqnarray}
where

\begin{eqnarray}\label{eq23}
{\mathcal R}_0 =\frac{16}{g}\frac{e\sigma_0}{{\overline T} c} \frac{\tau_{\varepsilon}}{\tau_{\bot}}\frac{H}{L},\qquad
 {\mathcal R}_0^V= \displaystyle \frac{32}{g}\frac{e}{{\overline T}c}\frac{\tau_{\varepsilon}}{\tau_{\bot}}.
\end{eqnarray}
According to Eq.~(23), the characteristic voltage responsivity 
${\mathcal R}_0^V$ does not explicitly  depend on the frequency of electron collisions $\nu$.
 
It is instructive that the responsivity at $V_G=0$ does not turn to zero because of the factor ${\mathcal F}(0) \neq 0$, so that $<\overline {\delta J_{\omega}} > 0$. 

\section*{Method and Results}

\begin{figure*}\centering
\includegraphics[width=18.0cm]{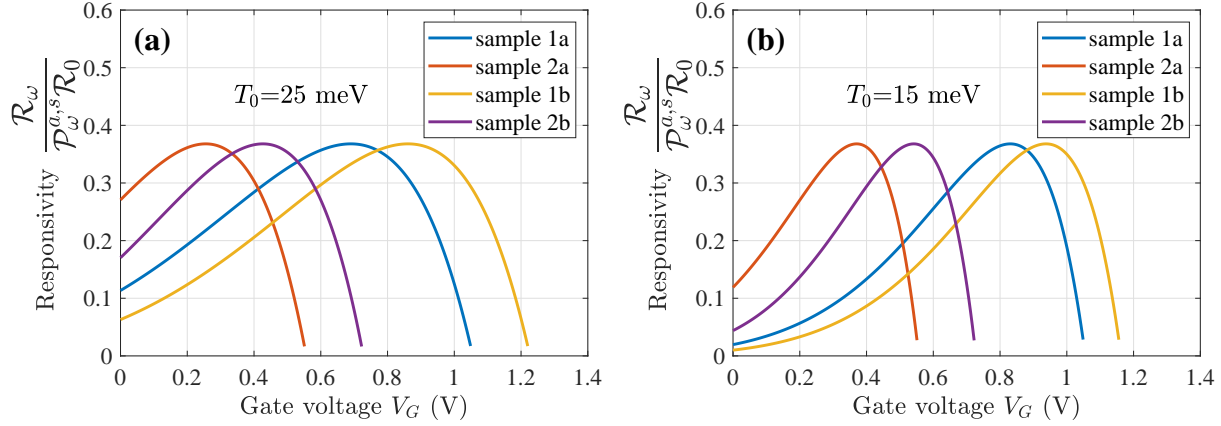}
\caption{
The normalized detector responsivity ${\mathcal R}_{\omega}/{\mathcal P}^{a,s}_{\omega}{\mathcal R}_0$  (the same for the asymmetric and symmetric THz radiation input)  as a function of the gate voltage $V_G$
for the GC-FETs 
with different band parameters:   (a) at $T_0 = 25$~meV and (b) $T_0 = 15$~meV.
} 
\label{F2}
\end{figure*}

Equations of the model were analyzed analytically and solved numerically. The resulting GC-FET characteristics - their responsivity found for different device samples are demonstrated in Figs.~2 - 5.

Figure~2 shows the normalized responsivity at the fundamental plasmonic resonance ${\mathcal R}_{\omega}/{\mathcal P}^a_{\omega}{\mathcal R}_0|_{\omega = \Omega}=  {\mathcal R}_{\omega}/{\mathcal P}^s_{\omega}{\mathcal R}_0||_{\omega = \Omega} ={\mathcal F}(V_G)$  (as a function of the gate voltage $V_G$)
for the devices with different $\Delta_C$, $\Delta_V$, $\Delta_M$, and the GC doping corresponding to the BL  flat band at $V_G = 0$ calculated using Eqs.~(5) and (20). In this case, the thermionic activation energy $\Delta_C - \mu_D = \Delta_V$. Equations~(5) and (20) are supplemented 
by the following relation for $\mu$  accounting for the effect of quantum capacitance~\cite{48,49,50,51}:

 \begin{eqnarray}\label{eq24}
(\mu- \mu_D)(\mu +\mu_D - 2\mu_0) = 2\mu_0eV_G,
\end{eqnarray}
where $\mu_0 = (\kappa_G\hbar^2v_W^2/4e^2W)$. For small (moderate) voltages, Eq.~(23) yields 
\begin{eqnarray}\label{eq25}
\mu \simeq  \mu_D + \frac{\mu_0}{(\mu_D+\mu_0}eV_G.
\end{eqnarray}

As seen from Fig.~2, the normalized responsivity, which might be rather high at $V_G = 0$,
exhibits  a maximum at a certain voltage $V_G^{max}$. The latter is different for different samples  depending on the device band parameters. A decrease in the temperature $T_0$ leads to somewhat sharper responsivity versus gate voltage dependence. This is associated with the specifics of the rectified current-voltage dependence  given by Eq.~(4). The fact that the maximum of function ${\mathcal F}(V_G)$ 
height is independent of $T_0$ is reflected in the dependences shown in Fig.~2.

As follows from Eq.~(18) and (22), the maximal values of ${\mathcal R}_{\omega}$ and ${\mathcal R}_{\omega}^V$ as functions of the signal frequency $\omega$ are reached at the  plasmonic resonances $\omega = \sqrt{n^2\Omega^2 - \nu^2} \simeq n\Omega$ for the asymmetrical input,
and $\omega = \sqrt{2n-1)^2\Omega^2/4 - \nu^2} \simeq (2n-1)\Omega/2$ for the symmetrical input,
where $n =1, 2,3,...$ is the plasmonic resonance index. At the fundamental resonances, ${\mathcal P}^a_{\omega}|_{\omega = \Omega}\simeq  2$ and 
${\mathcal P}^s_{\omega}|_{\omega = \Omega/2} \simeq  1$.

Figures~3 and 4 show the frequency dependence of the plasmonic oscillations factors ${\mathcal P}^a_{\omega}$ and ${\mathcal P}^s_{\omega}$ calculated
for different values of the plasmonic frequencies $\Omega$ and collision frequencies $\nu $.
According to Eq.~(22), these factors determine (proportional to) the spectral characteristics of the GC-FET detector responsivity.
 To account for the  electron collisions and the effect of their  viscosity on the plasmon damping, we
set~\cite{42} $\nu = \nu_{coll} + \nu_{visc}(\omega/\Omega)^2$, assuming
$\nu_{coll} = (1 - 2)$~ps$^{-1}$ and $\nu_{visc} = 0.25 $~ps$^{-1}$. In the GC-FETs with $L = (0.5 - 1.0)~\mu$m,  the latter corresponds to the electron viscosity     
$h \simeq (250 - 1000)$~cm$^2$/s that is in line with the observed values~\cite{42}.

In particular, Figs.~3 and 4 demonstrate that [in line with Eqs.~(17) and (22)] the responsivity exhibits  fairly sharp (resonant) maxima  at $\omega \simeq n \Omega$ and $\omega \simeq (2n-1)\Omega/2$ when 
 $\nu_{coll} = (1 - 2)$~ps$^{-1}$. 
 
Although  the GC-FETs with different methods of the THz radiation input 
exhibit the resonant response, the pattern of the spectral characteristics
shown in these plots are rather distinct, and  the resonance frequencies differ. This is associated with the excitation of different plasmonic modes (with different spatial distributions of the ac potential) using asymmetric and symmetric input. 
As seen, the amplitude of the plasmonic factor  maxima increases with increasing resonance index despite the strengthening of the viscosity effect.	 This is attributed to an increase in  the average AC electric field when   the number of its semi-periods, i.e.,  the index $n$ increase.

Figure~5 shows the dependences of the GC-FET detector current responsivity ${\mathcal R}_{\omega}$ corresponding
to the plasmonic factors of Figs.~3(b) and 4(b) calculated for $\nu_{coll} = 1$~ps$^{-1}$ and $\nu_{coll} = 2$~ps$^{-1}$ (solid lines). 
These dependencies exhibit  pronounced plasmonic resonances.
Since the responsivity ${\mathcal R}_{\omega}\propto \sigma_0{\mathcal P}_{\omega}
\propto {\mathcal P}_{\omega}/\nu$, the heights of the responsivity peaks for a larger index $n$ and for
a larger collisional frequency  $\nu$ are smaller. 
It is instructive that the voltage responsivity ${\mathcal R}^V_{\omega}$ for different collisional frequencies exhibits different behavior (at the chosen
load resistance, which is assumed to be inversely proportional to $\sigma_0$). 
However, as seen from Fig.~5 (dashed lines), the plasmonic resonances in the cases of much stronger
electron scattering ($\nu_{coll} = 4$~ps$^{-1}$ and $\nu_{coll} = 6$~ps$^{-1}$),
are substantially smeared.

Using Eq.~(14) and setting $\mu = 120 -140$~meV, $\kappa = 4$, $W = 10$~nm, and $L = (1- 2)~\mu$m, we obtain the following estimate: $\Omega/2\pi\simeq (0.53 - 1.14)$~THz.

Considering that the escape of a hot electron with the energy $\varepsilon \gtrsim \Delta_C$ from the GC over the BL is possible due to its scattering on an acoustic phonon.

 We might assume that the electron escape time $\tau_{\bot} \gg \tau_{ac}$, where $\tau_{ac}$ is the momentum relaxation time for
the electrons with the energy $\varepsilon \gtrsim \Delta_C$.
 The quantity $\tau_{ac}$  can be estimated as~\cite{52,53,54} $\tau_{ac} \simeq  1$~ps. 
Considering this, for rough estimates at the values $\Delta_C$ considered above we set $\tau_{\bot} \sim 10 - 20$~ps.
The electron energy relaxation time due to the interaction with the GC optical phonons is estimated as 
$\tau_{\varepsilon} \simeq 32- 65$~ps (compare, for example, with~\cite{32,34,55}). The
fast decay of optical phonons and the  interaction of the electrons in the GC with the interface optical phonons can lead to a decrease in $\tau_{\varepsilon}$ to the values about 10- 20~ps.
Setting $\mu = 140$~meV, $\nu = 1$~ps$^{-1}$,  $\tau_{\bot} = 10$~ps, $\tau_{\varepsilon} = 10$~ps,  $g= 1.64$, $H = 2L$,
and $(\Delta_C -\mu)/T_0 = 1$, we arrive at ${\mathcal R}_0 \simeq  4.1\times 10^2$ ~A/W.
This yields the characteristic voltage responsivity 
${\mathcal R}_{0}^V  \simeq 3.7\times  10^4$~V/W. The latter values are close
to the GC-FET current and voltage responsivities, ${\mathcal R}_{\omega}|_{\omega=\Omega}$ and 
${\mathcal R}_{\omega}^V|_{\omega=\Omega}$,
 at the plasmonic resonances.

\begin{figure*}\centering
\includegraphics[width=18.0cm]{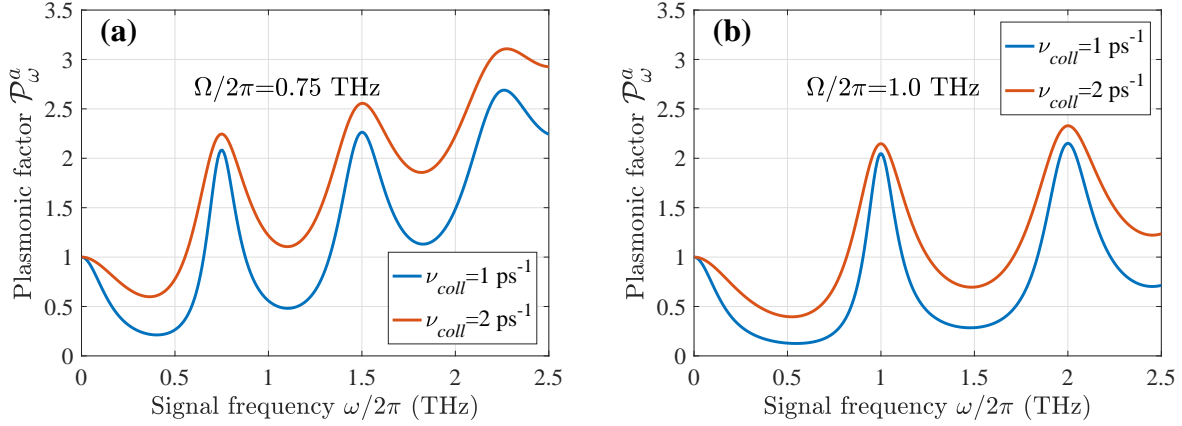}
\caption{
The plasmonic oscillation factor ${\mathcal P}^a_{\omega}$ of the GC-FETs with the asymmetric THz radiation input versus signal frequency $\omega/2\pi$: (a) for $ \Omega/2\pi = 0.75$~THz and (b) $ \Omega/2\pi = 1.0$~THz  ($\nu_{coll} = 1$~ps$^{-1}$ and 2~ps$^{-1}$ with $\nu_{visc} = 0.25$~ps$^{-1}$).
} 
\label{F3}
\end{figure*}

\begin{figure*}\centering
\includegraphics[width=18.0cm]{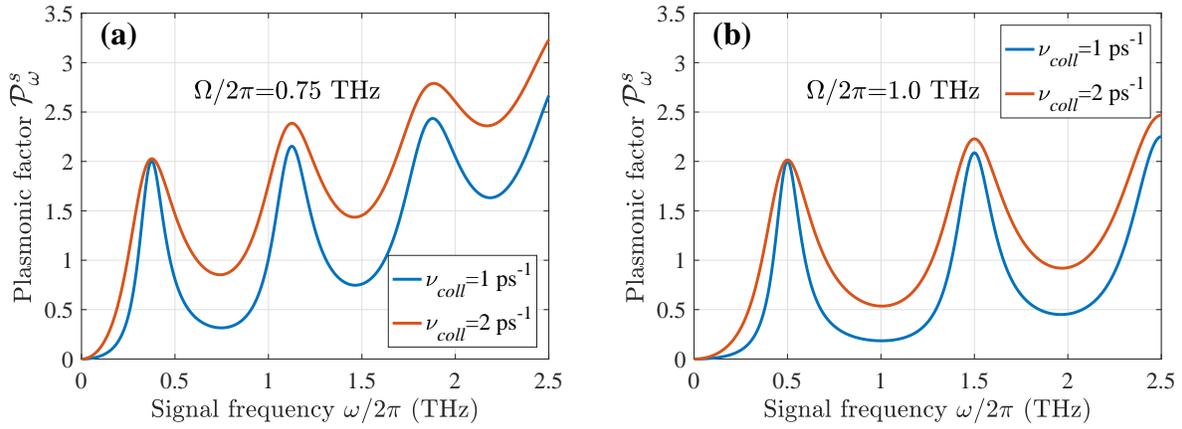}
\caption{
The same as in Fig.~3 but for the plasmonic oscillation factor
${\mathcal P}^s_{\omega}$ of
GC-FETs with the symmetric THz radiation input.
} 
\label{F4}
\end{figure*}

\begin{figure*}\centering
\includegraphics[width=18.0cm]{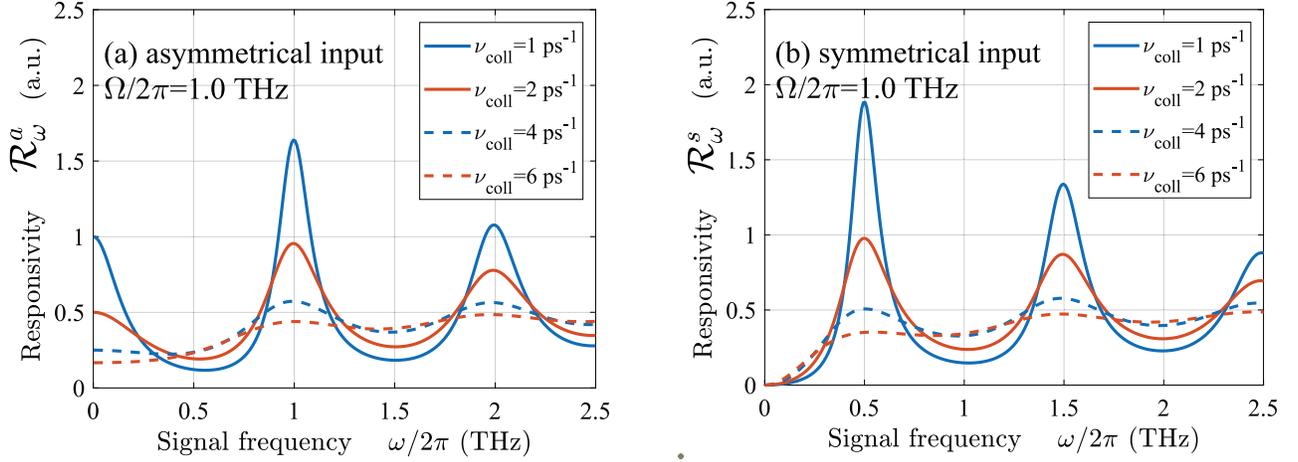}
\caption{
The  spectral characteristics of GC-FET detectors current responsivity with (a) asymmetric
and (b) symmetric THz radiation input ($\Omega/2\pi = 1.0$~THz) with different collision frequencies $\nu_{coll}$.
} 
\label{F5}
\end{figure*}

\section*{Discussion}

In Eq.~(10), which governs the electron energy balance in the GC, we
disregarded the electron cooling effect, associated with
thermionic emission. This effect can be accounted for by replacing the quantity $(\tau_{\varepsilon}/\tau_{\bot}){\mathcal F}(V_G)$
in Eq.~(22) by the factor $(\tau_{\varepsilon}/\tau_{\bot}){\mathcal F}(V_G)/[1+(\tau_{\varepsilon}/\tau_{\bot}){\mathcal F}(V_G)]$.
The pertinent distinction is small if $\tau_{\varepsilon} \lesssim\tau_{\bot}$.

 Using the relation between the GC mobility $M$ and the Fermi energy $\mu$:
$ M = ev^2_W/\nu\mu_D$, where the quantity $\mu_D/v_W^2$ is the electron fictitious mass, 
we find that the values of $\nu_{coll} = (1 - 2)$~ps$^{-1}$ and 
$\mu_D =(120 -140)$~meV assumed above  correspond to $M \simeq (4.3 - 7.1)\times 10^4$~cm$^2$/Vs. For  $\nu_{call} = (4 - 6)$~ps$^{-1}$ (see the dashed lines in Fig.5), one obtains $M \simeq (0.7 - 1.8)\times 10^4$~cm$^2$/Vs, 
which are  realistic GC mobilities at room or somewhat lower temperatures~\cite{53,55}. 
The electron mobility of the GC on b-P studied several years ago~\cite{56}, at $T_0 = (15 - 25)$~meV ($T_0 \simeq (180 -300)$~K) reaches the values $M\simeq (8 -9)\times 10^3$cm$^2$/Vs. This corresponds to $\nu_{coll} \simeq (8  - 10) $~ps$^{-1}$.  
Further improvements in the GC/b-P interface quality  or/and  
using the GC remote doping~(see, for example,~\cite{57,58}), one can reduce $\nu_{coll}$ increasing the plasmonic 
resonance  sharpness. Another option to decrease $\nu_{coll}$ is to use the positively biased back gate, which can electrically induce a sufficient electron density in the GC and, hence,
a proper value of the electron Fermi energy, eliminating the necessity of GC doping.
The plasmonic resonances and, hence, the pronounced resonant
response of the GC-FET detectors  might be more pronounced  for  larger plasma frequency
$\Omega$, i.e., in the devices with shorter GCs (smaller length $2L)$.  
In particular, if $2L = 0.5~\mu$m, the plasma oscillations quality factor   is about 8.2 even at $\nu_{coll} = 10$~ps$^{-1}$.
One needs to note that even at relatively high values of $\nu_{coll}$, the overdamped plasmonic oscillations  can provide elevated GC-FET detector responsivities despite the resonant peaks vanishing. 

Electron thermal conductivity along the GC~\cite{59}, which 
leads to the  transfer of a portion of the electron heat to the side contacts can
reduce the electron temperature  and
 smooth down  the spatial nonuniformities of the electron density. The latter  
can particularly affect the resonant maxima height with increasing plasmonic mode index $n$.

Fairly high values of the GC-FET responsivity are due to the long electron energy relaxation time $\tau_{\varepsilon}$ inherent for GCs. However, the speed of the photodetectors using the hot electron  bolometric mechanism is limited by
the inverse electron energy relaxation time $\tau_{\varepsilon}^{-1}$ (see, for example,~\cite{34,54}).
This implies that the operation of the THz GC-FET detectors under consideration 
(with the parameters used in the above estimates) might be limited to the modulation frequencies in  the GHz range.

The GC-FET detector dark current limited detectivity $D^{*}_{\omega} ={\mathcal R}_{\omega}/\sqrt{2e {\overline j}} $ (see, for example,~\cite{60})  depends, in particular, on the dark current density. 
As  follows from Eqs.~(2) and (25), the dark current density is

\begin{eqnarray}\label{eq26}
{\overline j}\simeq  j^{max} \exp\biggl(-\frac{\Delta_M}{T_0}\biggr)\exp\biggl[\frac{\mu_0}{{(\mu_D + \mu_0)}}\frac{eV_G}{T_0}\biggr]\bigg[1 - \exp\biggl(-\frac{eV_G}{T_0}\biggr)\biggr].
\end{eqnarray}
For  low gate voltages ($eV_G < T_0$), the latter
tends
 to zero as  ${\overline j} \propto V_G$.
 Since the responsivity in the limit of small $V_G$ is a constant (see Fig.~2), this implies that  the GC-FET detector  detectivity as a function of the gate voltage  increases with decreasing $V_G$ as
 
 \begin{eqnarray}\label{eq27}
 D^{*}_{\omega} \propto \frac{1}{\sqrt{V_G}}. 
\end{eqnarray}
This also means that at low values of $V_G$, the GC-FET noises might be determined by
other mechanisms (not by the dark current).

If the electron interactions at the GC/b-AsP interface are relatively  strong
(leading to high values of $\nu_{coll}$ and 
preventing the pronounced plasmonic resonance) is a critical issue, the GC-FET structure can be modified by using the length of the gate and the b-AsP  layer
markedly smaller than the length of the GC $2L$.

In principle, the GC-FET detectors can be based on the p-type GC, in which $\mu_D <0$. This might exhibit
advantages associated with smaller $\Delta_V$ compared to $\Delta_C$ (see Table 1).
In the detectors with the p-type GC, a proper thermionic activation energy
$\Delta_V +\mu_D$ can be achieved at smaller $\mu_D$, i.e., at the  lower carrier (hole) densities. 
However, the  adequate BL and metal gate  band alignment 
might be a problem. This problem can be avoided in the GC-FET structures, in which
both the GC and the gate are made of  p-type  graphene layers (double-GC-FETs).
Such  double-GC-FETs can exhibit markedly different
plasmonic properties. This is because of the possible plasmonic response of the carriers in the double-GG (see, for example,~\cite{61,62,63,64,65,66,67}). The plasmonic  response in the double-GC-FETs depends on the contacts.
Depending on the geometry of these contacts, the plasmonic factor can be a fairly
different function of the signal frequency. The bolometric detectors based on the double-GC-FET
structures with the barrier b-As$_x$P$_{1-x}$ are beyond the scope
of our present study and  require a separate treatment.

\section*{Conclusions}
We proposed and evaluated  the THz graphene-channel-
FET detectors  with  the black-Arsenic,   black-Phosphorus,  or black-Arsenic-Phosphorus
barrier gate  layers.
The operation of these detectors is associated with the hot carrier bolometric effect, i.e., with the carrier heating
by incoming THz radiation, causing their thermionic emission from the graphene channel into the gate. 
Such a THz  GC-FET detector can exhibit fairly high characteristics.
The excitation of plasmonic oscillations in the graphene channel leads to a strong resonant enhancement of the detector responsivity and detectivity.

The realization of the proposed GC-FET bolometric detectors 
with elevated characteristics 
is enabled by the effective carrier heating in graphene accompanied by the effective plasmonic oscillation excitation and the possibility of a proper
band alignment between the graphene  channel and the barrier layer.

\section*{Acknowledgments}
Financial support is provided by The Japan Society for Promotion of Science (KAKENHI Grants $\#$ 20K20349 and $\#$ 21H04546), Japan; 
RIEC Nation-Wide Collaborative Research Project $\#$ R04/A10; 
and by AFOSR (contract number FA9550-19-1-0355.

\section*{Author contributions statement}

V.R. conceived the device concept and developed its model, C.T. and M.R. conducted the calculations and presented the results, T. O., V. M, and M.S.
analyzed the model, obtained results, and provided financial support. 
 All authors reviewed the manuscript. 

\section*{Data availability}
All data generated or analyzed during this study are included in this published article.

\section*{Additional information}
\textbf{Competing interests:} The authors declare that they have no competing interests. 


\end{document}